\documentstyle[12pt,epsfig]{article}
\def\setb@se#1{\baselineskip=#1 \normalbaselineskip=#1}
\setb@se{14pt}
\newcommand{\be}{\begin{equation}}
\newcommand{\ee}{\end{equation}}

\newcommand{\vp}{\varphi}
\newcommand{\vt}{\vartheta}

\newcommand{\tr}{{\rm tr}}

\newcommand{\const}{\mbox{\rm const.}}

\newcommand{\p}{\partial}

\begin{document}

\begin{center}
{\Large Internal structure of Einstein-Yang-Mills-Dilaton black holes }

\vspace{2 mm}

{\sl O. Sarbach, N. Straumann, and M. S. Volkov }

{\it Institute for Theoretical Physics, University of Zurich-Irchel, 
Winterthurerstrasse 190, 
CH--8057 Zurich, Switzerland}

\end{center}

\noindent
{\sl We study the interior structure of the 
Einstein-Yang-Mills-Dilaton black holes as a function of
the dilaton coupling constant $\gamma\in [0,1]$. 
For $\gamma\neq 0$ the solutions have no
internal Cauchy horizons and the field amplitudes follow 
a power law behavior
near the singularity. As $\gamma$ decreases, the solutions develop more and
more oscillation cycles in the interior region, whose number becomes 
infinite in the limit $\gamma\rightarrow 0$. }

\vspace{10 mm}

\noindent
{\bf I.} The much discussed mass inflation scenario \cite{IP}
provides a picture of what happens to the unstable Cauchy horizons 
of the Kerr-Newman family when the back reaction of the 
blue-shifted perturbations is taken into account. In this scenario
the blue-shifted influx, due to the inavoidable gravitational wave tail
of a realistic gravitational collapse, causes the local mass function 
to blow up exponentially near the Cauchy horizon. Thereby, 
a singularity is produced along the Cauchy horizon, which leads to infinite
tidal forces on a free-falling object. This singularity is, however,
lightlike and mild, accompanied with a divergence of the conformal
curvature, but with bounded shear and expansion. The genericity
of the mass inflation picture has been established in a series of analytical
and numerical studies of a class of vacuum and electrovacuum solutions
\cite{minfl}. 

What really happens very close to the singularity along the horizon 
cannot be decided by classical physics. This is presumably not even a 
physically relevant question, because the black hole will have 
evaporated or merged with other black holes in a cosmological big crunch. 
It may, nevertheless, be of some interest to study the interior black hole
solutions for classical field--theoretical matter models which are
important in high-energy particle physics. 

Such investigations were initiated in ref. \cite{dima}, 
where it was shown that for pure Yang-Mills fields a new type
of infinitely oscillating behavior with exponentially growing
amplitude is developed. It is natural to ask, whether this cyclic
repetition with associated mass inflation is generic when other
matter fields are included. In \cite{BLM}, \cite{dima1}
it was shown that the interior solution changes
qualitatively when a Higgs field is included. 

In the present contribution we discuss the interior structure of 
black holes for the Einstein-Yang-Mills-Dilaton (EYMD) system,
considering the dilaton coupling constant, $\gamma$, as a
bifurcation parameter of the associated dynamical system. 
It turns out that for $\gamma=1$ the behavior is drastically 
different from that of the EYM system ($\gamma=0$), 
in that there are no oscillations; this phenomenon has also been
observed in \cite{dima2}. All fields evolve quite monotonically
when the radial Schwarzschild coordinate $r$ (internal time)
approaches 0. This has prompted us to study the change of the 
phase portrait as $\gamma$ decreases from 1 to 0. 
Our analytical and numerical results show quite convincingly,
that with decreasing $\gamma$ more and more cycles are 
developing, and that the limit $\gamma\rightarrow 0$ is quite
singular. For $\gamma\neq 0$ the interior solutions have no 
Cauchy horizons. We believe that this is true quite generically 
for a large class of nonlinear field--theoretical matter models. 
Further work on this issue is in progress.

\newpage

\noindent
{\bf II.}
The black hole solutions we shall be considering arise within the context
of the SU(2) EYMD theory, whose action reads in standard notations
\begin{equation}
\label{eq:action}
{\cal S}=\int
\left(-\frac{1}{4}R+\frac{1}{2}\, \p_\mu\Phi\, \p^\mu\Phi
+\frac{1}{2}e^{2\gamma\Phi}
\tr\, F_{\mu\nu}F^{\mu\nu}\right) \sqrt{-g} \,d^4 x. 
\end{equation}
The dilaton coupling $\gamma$ is assumed to have values 
in the interval $[0,1]$. We are especially interested in the limiting
behavior for $\gamma\rightarrow 0$. 
When $\gamma$ vanishes, the dilaton decouples, in which case
one can put $\Phi=0$.
In the static spherically symmetric case the metric is given by 
\begin{equation}                                                           \label{1}
ds^2=S^2 N dt^2-\frac{dr^2}{N}-r^2 (d\vt^2+\sin^2 d\vp^2)
\end{equation}
and the gauge potential $A$ for a purely magnetic YM field 
can be parametrized as
\begin{equation}                                                         \label{2:0}
A=w(r)(-{\rm T}_{2}\, d\theta+{\rm T}_{1}\sin\theta\, d\varphi)+
{\rm T}_{3}\cos\theta\, d\varphi, 
\end{equation}
where the group generators ${\rm T}_a$ are chosen as $\tau^a/2i$
($\tau^a$=Pauli matrices). 
The functions $w$, $S$, $N$ and the dilaton $\Phi$ depend only 
on $r$. It will be sometimes  convenient to express $N$ in terms 
of the mass function $m$ as $N=1-2m/r$. 
The field equations, corresponding to (\ref{eq:action}) read
\[
(rN)^{\prime }+r^{2}N\Phi ^{\prime \,\,2}+U=1,
\]
\[
\left( S Nr^{2}\Phi ^{\prime }\right) ^{\prime
}=\gamma S U,
\]
\[
r^2\left( NS e^{2\gamma\Phi }\,w^{\prime }\right) ^{\prime }=
S e^{2\gamma\Phi 
}\,
w(w^{2}-1),
\]
\begin{equation}                                      \label{4}
S ^{\prime }=S\, (r\phi ^{\prime \,\,2}+ 2e^{2\gamma\Phi
}\,w^{\prime\, 2}/r), 
\end{equation}
where $U=2e^{2\gamma\Phi }\left( Nw^{\prime\, 2}
+(w^{2}-1)^{2}/2r^{2}\right)$.
These equations are invariant under the scale transformations
 \begin{equation}                                           \label{4:1}
r\mapsto e^\lambda r,\  \ \  N\mapsto N,\ \ \  S\mapsto e^{-\lambda}S,\ \ \ 
\Phi\mapsto\Phi+\frac{\lambda}{\gamma},\ \ \  w\mapsto w. 
 \end{equation}
The function $S$ can be eliminated from the system (\ref{4}). 
When $N$, $\Phi$ and $w$ are known,  $S$ can be expressed as
\begin{equation}
S=\exp \left(-\int_{r}^\infty (r{\Phi'}^2+\frac{2}{r}e^{2\gamma\Phi}{w'}^2)dr\right). 
\end{equation}

For any $\gamma\in [0,1]$, equations (\ref{4})
are known to possess a family of black hole 
solutions \cite{dima3}, \cite{LM}. Close to the event horizon they behave
as follows:
 \be                                                  \label{5}
   N= (1-V_h) x + O(x^{2}),\ \ 
\Phi = \Phi_h+\frac{\gamma V_h}{(1-V_h)} x
+ O(x^2), \ \ 
   w = w_h+
\frac{w_h(w_h^2-1)}{(1-V_h)}x
 + O(x^{2}),
 \ee
where $x=(r-r_h)/r_h$, 
$V_h=e^{2\gamma\Phi_h}(w_h^2-1)^2/r_h^2$, 
and the scaling symmetry (\ref{4:1}) can be used to set $\Phi_h=0$. 
Note that at the event horizon one has $N'(r_h)>0$.
The asymptotic behavior at infinity is described by
\be                                        \label{6}
 N = 1-\frac{2M}{r}+\frac{D^2}{r^2} + O(\frac{1}{r^3}),
\ \ \ \Phi= \Phi_\infty-\frac{D}{r} + O(\frac{1}{r^2}),\ \ \
w = \pm \left(1-\frac{c}{r}\right)
+ O(\frac{1}{r^2}),
 \end{equation}
where $M$ is the ADM mass and $D$ is the dilaton charge. 
The black hole solutions 
with these asymptotics are numerically known in the whole interval
$r_h\leq r <\infty$ \cite{dima3}, \cite{LM}. Their behavior 
is qualitatively similar for all values of $\gamma$: the functions
$N$ and $\Phi$ are monotone in the exterior region, interpolating 
between the boundary values given by Eqs. (\ref{5}), (\ref{6}),
whereas $w$ oscillates within finite bounds. 
The solutions form
a 2-parameter family labeled by $r_h$, and
$n$, the number of nodes of $w$ in $[r_h,\infty)$.  
In what follows we shall consider the extension
of these solutions into the interior region of the black hole, 
$0\leq r< r_h$. It will turn out that the behavior of the 
solutions in the interior region changes dramatically 
with varying $\gamma$. 

\noindent
{\bf III.}
When extending the black hole solutions into the
interior region, the following phenomenon is observed:
For $\gamma=0$ (the dilaton is absent) 
the interior solution is characterized by  violent oscillations 
of the metric coefficients and the gauge field strength \cite{dima}.  
The amplitude of these oscillations tends to infinity as the system approaches
the singularity. For $\gamma=1$ the behavior is rather different, 
without any peculiar variations (see Fig.1). 
It is interesting to study more closely 
what happens when $\gamma$ decreases form 1 to 0. 
Our analytical and numerical investigations lead to the 
following qualitative picture:
For small values of $\gamma$
the solution in the vicinity of the horizon follows closely that for 
$\gamma=0$. 
The smaller the $\gamma$, the more oscillation cycles
the solution exibits. However, as long as $\gamma\neq 0$, 
the functions assume a power law behavior
in the vicinity of $r=0$: 
\begin{equation}                                   \label{123}
   N \propto -N_1 r^{-(1+p^2)},\ \ \ 
\Phi \propto \Phi_1 + p\ln(r),\ \ \ 
   w \propto w_0-b r^{2(1-\gamma p)},
 \end{equation}
with $b,N_1,p>0$, and the values of $p$ are restricted by
\begin{equation}
\label{eq:pc}
\sqrt{\gamma^2+1}-\gamma<p<\frac{1}{\gamma}. 
\end{equation}

\begin{figure}
\hbox to\hsize{
  \epsfig{file=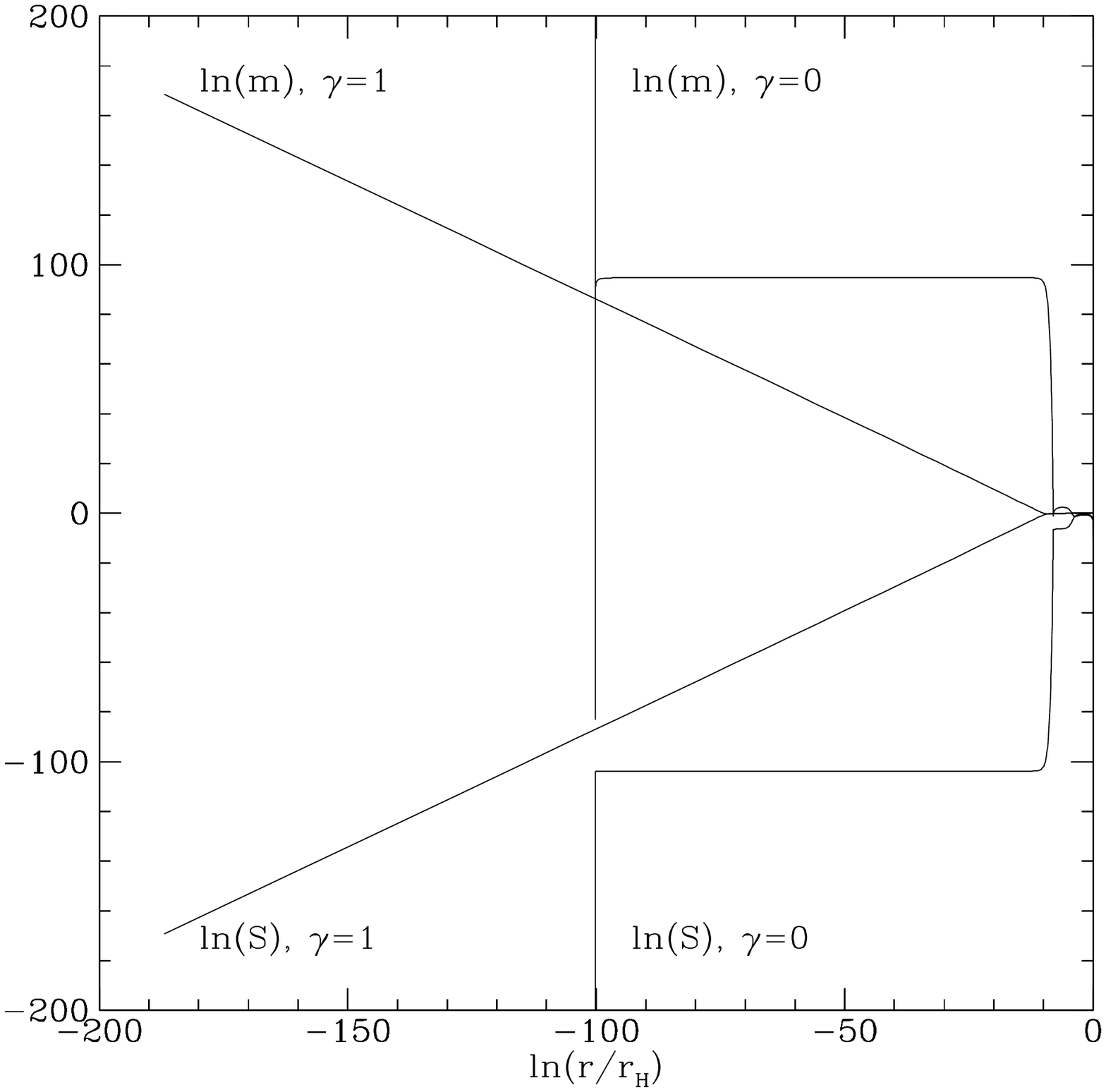,width=0.48\hsize,%
      bbllx=1.8cm,bblly=5.5cm,bburx=20.0cm,bbury=20.0cm}\hss
  \epsfig{file=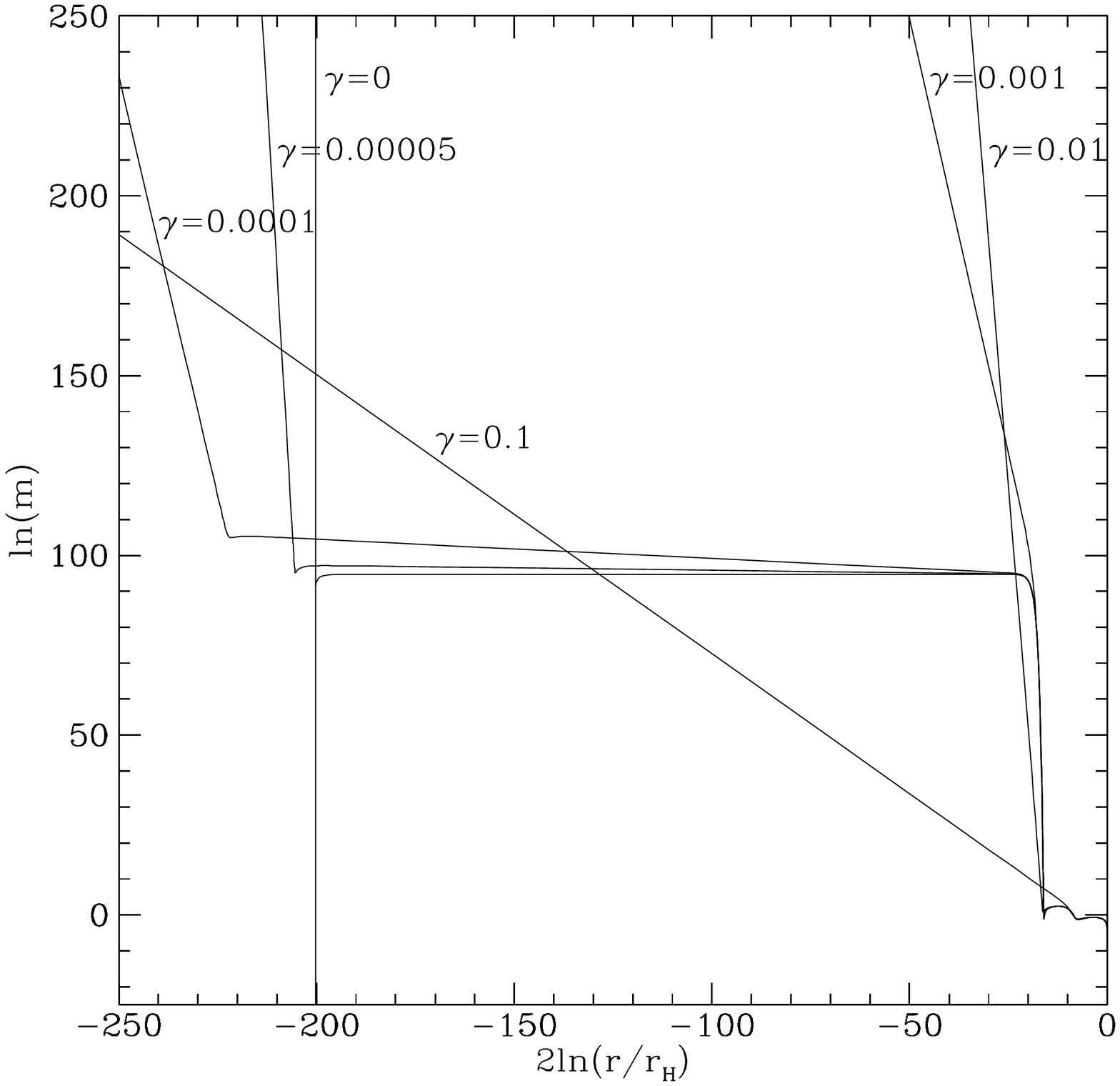,width=0.48\hsize,%
      bbllx=1.8cm,bblly=5.5cm,bburx=20.0cm,bbury=20.0cm}
  }
\caption{\it The interior solutions for the $n=1$, $r_h=2$ EYMD black holes. 
On the left: for $\gamma=0$ the metric functions $m$ and $S$
can be approximated by step-functions
whose amplitudes grow exponentially as $r\rightarrow 0$. 
For $\gamma=1$,  $m$ and $S$ are described asymptotically by linear functions
-- on a logarithmic scale. The picture on the right
shows how the steps transform into straight lines as $\gamma$
increases.
}
\end{figure}

In order to gain some qualitative understanding of this phenomenon, 
it is helpful to introduce the new variables
 \begin{equation}
x=w' e^{\gamma\Phi},\ \ \ 
y=-\frac{(w^2-1)^2}{r^2 N}e^{2\gamma\Phi},\ \ \
z=r\Phi'. 
 \end{equation}
In addition, the following approximations in the differential equations
are numerically justified for the interior solutions: 
\begin{itemize}
\item{$w\simeq\const$ (but $w'$ is kept),}
\item{$e^{2\gamma\Phi} (w^2-1)^2 \gg r^2$, }
\item{the term 
$w(w^2-1)/r^2$ in Eqs.(\ref{4}) can be neglected. }
\end{itemize}
This yields, as an approximate first integral of the field equations, 
\begin{equation}
SNe^{2\gamma\Phi}w'\simeq\const
\end{equation}
Eqs. (\ref{4}) can then be truncated to the 
following dynamical system: 
$$
\dot{x} = x(y+\gamma z -1), 
$$
$$
\dot{y} = -y(2x^2-y-1+z^2+2\gamma z), 
$$
\begin{equation}                            \label{ds}
\dot{z} = -\gamma(2x^2-y)+yz, 
 \end{equation}
where a dot stands for differentiation with respect to $t=-\ln(r)$. 
We are interested in the behavior of the solutions as $t\rightarrow \infty$. 
First of all, we note that the planes $\{ x=0 \}$ and  $\{ y=0 \}$ 
are invariant sets and hence divide the phase space into four regions. 
Since $N$ is negative and the equations are symmetric under 
$x\mapsto -x$, we restrict ourselves to the invariant region $x>0$, $y>0$.
 The critical points in this region are:
\begin{figure}
\hbox to\hsize{
  \epsfig{file=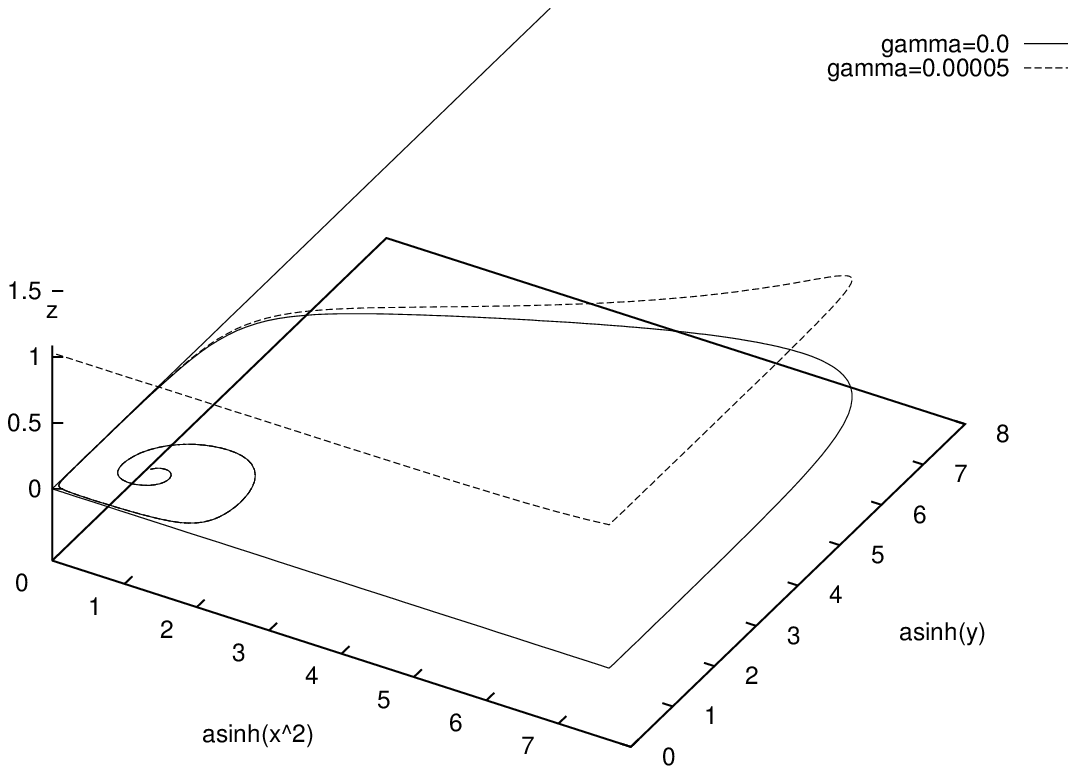,width=0.5\textwidth,height=6cm}\hss
  \epsfig{file=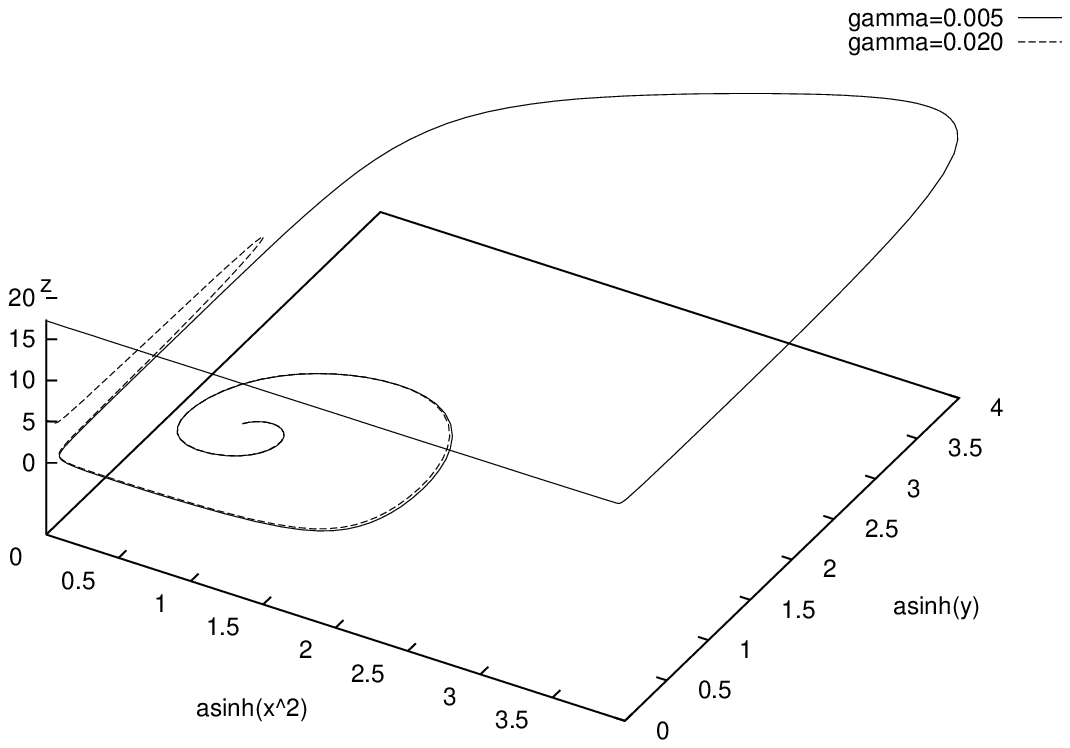,width=0.5\textwidth,height=6cm}\hss
  }
\caption{\it On the left: solutions of the dynamical system (13) for 
$\gamma=0$ and $\gamma=0.00005$. 
For $\gamma=0$ the trajectory is bound to stay
in the $z=0$ plane always spiralling around the center, 
whereas for 
$\gamma=0.00005$ it leaves the plane
and converges to the $z$-axis. 
On the right: the solutions for $\gamma=0.005$ and $\gamma=0.020$.
For $\gamma=0.005$, one additional oscillation cycle is performed. 
}
\end{figure}

\begin{enumerate}

\item{$x=\alpha\beta$, $y=\alpha\beta$, $z=-\alpha\gamma$, where $\alpha=1/(1+2\gamma^2)$ and $\beta=1+\gamma^2$.}
\item{$x=0$, $y=0$, $z=p$, where $p$ is arbitrary.}

\end{enumerate}
The eigenvalues corresponding to the first critical point
are $\lambda_1=\alpha\beta$, $\lambda_{2,3}=\alpha\beta(1\pm i\sqrt{15+32\gamma})/2$, 
hence it is an attractive center in the limit $t\rightarrow -\infty$. 
The solutions starting at this point spiral outwards when $t$ increases. 
For $\gamma=0$ one has 
$z\equiv 0$ and Eqs.(\ref{ds}) reduce to the 2-dimensional system
 $$
\dot{x} = x(y-1),
$$
\begin{equation}
\dot{y} = y(-2x^2+y+1).
 \end{equation}
Now, for this system one can show that the spiraling motion
around the focal point can never stop. 
The trajectories are bound to spiral around the center $(1,1)$
for all $t$. At the same time, they have to remain
in the region $x>0$ and $y>0$. As a result, after each revolution
the trajectories come closer and closer to the second critical point,
$(0,0)$, which corresponds to the spacetime singularity,
but can never be reached. This explains qualitatively 
the unbounded oscillations
for the pure EYM case \cite{dima}. It is interesting to observe
that the critical point around which the spiralling occures
has no  physical meaning by itself. Indeed, it can only be
reached for $t\rightarrow -\infty$, which corresponds to
$r\rightarrow +\infty$, contradicting the assumption 
$r<r_h$. 

On the other hand, for $\gamma\neq 0$ the solutions can no longer stay in the 
plane $\{ z=0\}$ for all $t$. It turns out that the appearance of the 
additional degree
of freedom completely changes the behavior of the system. 
When $\gamma$ is small the 
trajectory can stay for a long time in the vicinity of the plane spiralling
outwards. The smaller $\gamma$, the smaller $\dot{z}$
and more revolutions are executed by the trajectory. 
As long as $p=z$ is small, the second critical point
is repulsive, since the corresponding eigenvalues are
$\lambda_1=1-\gamma p$, 
$\lambda_2=p^2+2\gamma p-1$, and $\lambda_3=0$. For small $p$ one has $\lambda_1>0>\lambda_2$. 
However, when $p=z$ becomes large enough to fulfill the condition (\ref{eq:pc}), 
one will have $\lambda_1>0$, $\lambda_2>0$, 
and so the critical point $(0,0,p)$ will become attractive. 
One can also show that it is stable, despite the zero eigenvalue. 
As a result, having performed only a finite number of oscillations,
the trajectory will eventually be attracted and end up at the singularity.
The linearized solution near the critical point reproduces the 
power law behavior given by Eq.  (\ref{123}). 

\noindent
{\bf IV.} The interior solutions decribed above do not 
have  inner horizons. It is natural to ask whether this property 
persists for all values of $n$ and $r_h$. 
For $\gamma=0$ some of the the solutions are known to display 
Cauchy horizons.  Specifically, for each $n>1$
there is one distinguished black hole solution with radius $r_h(n)$, 
which has an inner horizon at some $r=r_{-}(n)$ \cite{dima}. 
It turns out that for $\gamma\neq 0$ this can never happen. 
The proof of this statement is fairly simple
and goes as follows: When $\gamma\neq 0$, the scaling symmetry (\ref{4:1}) 
implies the existence of the conserved Noether current
\begin{equation}
\label{eq:current}
J=r^2\left(\frac{1}{2}SN'+NS'-NS\frac{\Phi'}{\gamma}\right). 
\end{equation}
The conservation condition $J'=0$ can be straightforwardly verified 
with the help of Eqs. (\ref{4}). 
Taking the limit $r\rightarrow r_h$, we obtain
\be
\label{eq:current1}
J= \frac{r^{2}_{h}}{2} S_h N'(r_h)>0, 
\end{equation}
since  $S(r)$ is everywhere positive, and $N'>0$ at the event horizon. 
Now, assuming that an inner horizon exists at some $r_{-}<r_h$, 
such that $N(r_{-})=0$, $N'(r_{-})<0$, and taking the limit 
$r\rightarrow r_{-}$, we arrive at 
\be
\label{eq:current2}
J= \frac{r^{2}_{-}}{2} S(r_{-}) N'(r_{-})<0, 
\end{equation}
which contradicts (\ref{eq:current1}). 
The addition of a dilaton field thus eliminates
the Cauchy horizons. We believe that the same is true for a large
class of non-linear matter models. 

A more detailed account of this work is given in the diploma thesis 
by OS \cite{dipl}. MSV thanks the organizers of 
``The Internal Structure of Black Holes and Spacetime Singularities''
workshop at Haifa for an interesting meeting and the generous hospitality.

\end{document}